\documentclass{ws-rv961x669}
\usepackage[square]{ws-rv-van}     % numbered citation/references (default)
\usepackage{ws-rv-thm}     % comment this line when `amsthm / theorem / ntheorem` package is used
\usepackage{subfigure}     % required only when side-by-side / subfigures are used
\makeindex
\begin{document}

\chapter[Recent Results from the ANTARES Neutrino Telescope] {Recent Results from the ANTARES Neutrino Telescope}\label{ra_ch1}

\author[P. Coyle and C. W. James ]{Paschal Coyle }
%\index[aindx]{Author, F.} % or \aindx{Author, F.}
%\index[aindx]{Author, S.} % or \aindx{Author, S.}

\address{Centre de Physique des Particules de Marseille,\\
163 Avenue de Luminy, Case 902, Marseille 13288 cedex 09, France \\
coyle@cppm.in2p3.fr\\ }

\author[P. Coyle and C. W. James ]{Clancy W. James }

\address{Friedrich-Alexander-Universit\"at Erlangen-N\"urnberg, \\
Erlangen Centre for Astroparticle Physics, \\
Erwin-Rommel-Str. 1, 91058 Erlangen, Germany \\
clancy.james@physik.uni-erlangen.de}

\author[]{for the ANTARES Collaboration }

\begin{abstract}
The ANTARES deep sea neutrino telescope has been taking data continuously since its completion in 2008.
With its excellent view of the Galactic plane and good angular resolution the telescope can 
constrain the origin of the diffuse astrophysical neutrino flux reported by IceCube.     
Assuming various spectral indices for the energy spectrum of neutrino emitters, 
the Southern sky and in particular central regions of our Galaxy have been studied searching for point-like objects, 
for extended regions of emission, and for signal from transient objects selected 
through multi-messenger observations. For the first time, cascade events are used for these searches.

ANTARES has also provided results on searches for hypothetical particles (such as magnetic monopoles
and nuclearites in the cosmic radiation), and multi-messenger studies of the sky in combination
with various detectors. Of particular note are the searches for dark matter: the limits obtained
for the spin-dependent WIMP-nucleon cross section surpassing those of current direct-detection
experiments.

\end{abstract}

%\markright{Customized Running Head for Odd Page} % default is Chapter Title.

\body

%\tableofcontents

\section{Introduction}\label{ra_sec1}

Being weakly interacting the neutrino is unique and complementary to other astrophysical probes such as 
multi-wavelength light and charged cosmic rays. 
The neutrino can escape from regions of dense matter or radiation fields 
and can travel cosmological distances without being absorbed. Neutrinos make it possible to distinguish unambiguously
between hadronic and electronic acceleration and to localise these acceleration sites more precisely than cosmic 
rays detectors. High-energy neutrinos may also be produced by the annihilation of dark matter particles which may have 
accumulated in the cores of massive astrophysical objects such as the Sun. Since the 
recent observation of a diffuse flux of cosmic neutrinos by the IceCube Collaboration, 
an understanding of its origin has become a top priority for the astroparticle physics community. 

The ANTARES neutrino telescope being located in the Northern Hemisphere has a good visibility for the Galactic 
plane and, due to the exceptional optical properties of the deep seawater, provides an excellent angular resolution on 
the neutrino direction. The ANTARES detector (see Ref. \cite{antares} for details), 
is located 40 km offshore from Toulon at a depth of 2475 m. It was completed on 29 May 2008, 
making it the largest neutrino telescope in the northern hemisphere and the first to operate in the deep sea. 
The telescope is optimised to detect upgoing high energy neutrinos ($>$ 10 GeV) by observing the Cherenkov light 
produced in seawater from secondary charged leptons that originate in neutrino interactions near 
the vicinity of the instrumented volume. Because of the long range of the muon, neutrino interaction vertices tens of 
kilometres away from the detector can be observed thereby increasing the effective volume. Other neutrino flavours 
are also detected, although with lower efficiency and degraded angular precision because of the shorter range of the 
corresponding leptons.

The detector infrastructure comprises 12 mooring lines hosting light sensors. Due to its location in the deep sea, the 
infrastructure also provides opportunities for innovative measurements in Earth and sea sciences (see for example Ref. \cite{Tamburini}).
Another project benefiting from the deep sea infrastructure is an R\&D system of hydrophones which investigates the 
detection of ultra-high energy neutrinos using the sound produced by their interaction in water. This system called 
AMADEUS (Antares Modules for the Acoustic Detection Under the Sea) \cite{amadeus} is a feasibility study 
for a prospective future large-scale acoustic detector. 

The decommissioning of the ANTARES telescope is planned for 2017, at which point the KM3NeT neutrino telescope 
\cite{km3net} will have surpassed ANTARES in sensitivity. 

\section{General Description of the detector}

\begin{figure}[hb]
\centerline{\includegraphics[width=10cm]{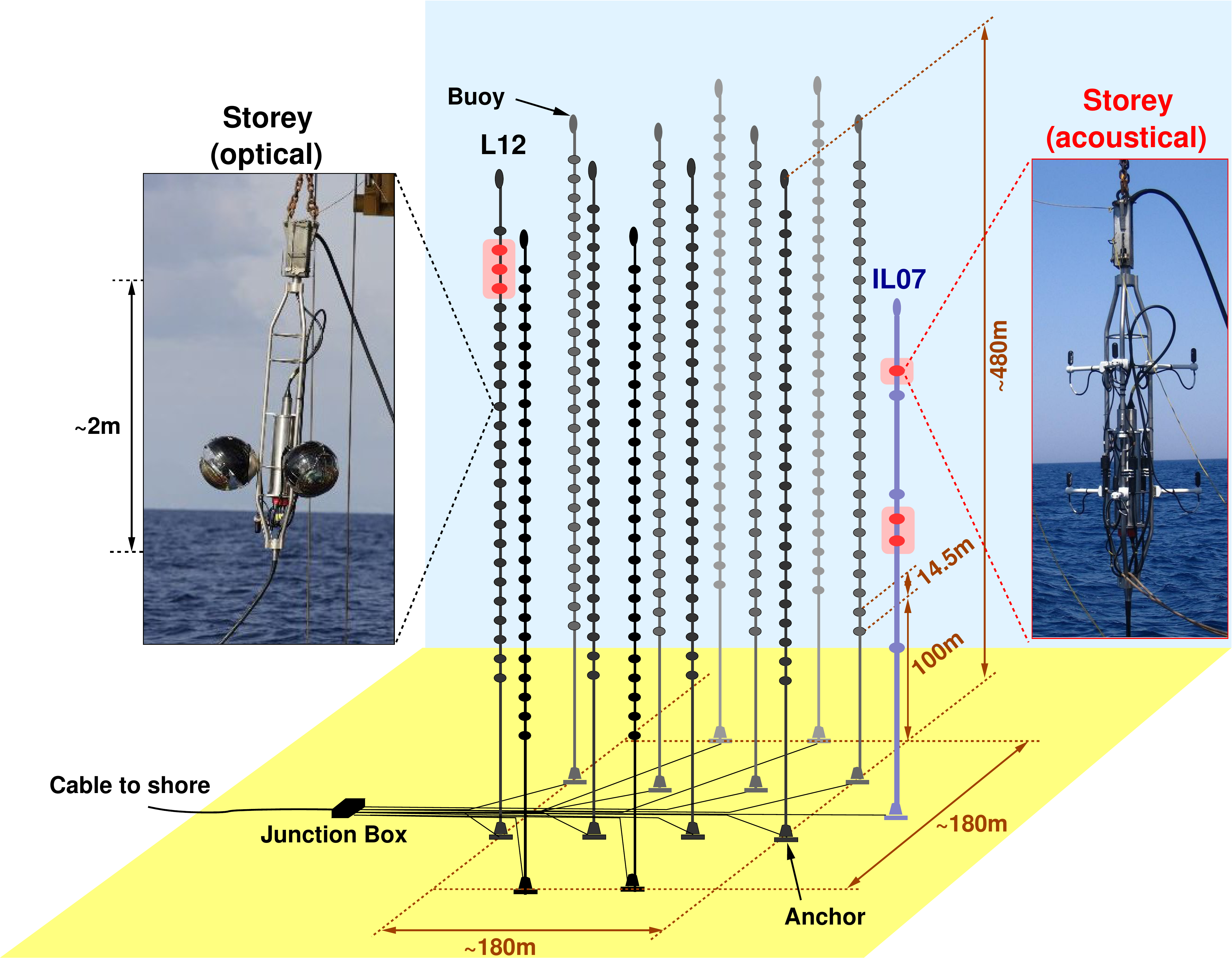}}

\caption{Schematic view of the ANTARES detector.}
\label{antares-view}
\end{figure}

A schematic view of the ANTARES neutrino telescope is given in \fref{antares-view}. 
The basic detection element is the Optical Module which houses a 10 inch photo-multiplier tube inside a 
pressure-resistant glass sphere.
Each node of the three-dimensional telescope array is the assembly of a mechanical 
structure, which supports three optical modules, looking downwards at $45^\circ$, and a titanium container 
housing the offshore electronics and embedded processors. 
In its nominal configuration, a detector line comprises 25 storeys linked together with an electro-mechanical cable. 
The distance between storeys is 14.5 m with the first storey starting 100 m from the seabed. The line is anchored on 
the seabed and is held vertical by a submerged buoy. The full neutrino telescope comprises 12 lines, 11 with the 
nominal configuration, the twelfth line being equipped with 20 storeys and completed by devices dedicated to 
acoustic detection. Thus, the total number of the OMs installed in the detector is 885.  The lines are arranged in an 
octagonal configuration, with a typical interline spacing of 60-70 m. The infrastructure is completed by the 
Instrumentation Line, which supports the instruments used to perform environmental measurements. 

The detection lines are flexible and variations in the intensity and direction of the sea current can induce a coherent displacement, 
typically of the order of a few meters for the uppermost storeys. In addition, the storeys may rotate about the vertical axis. 
For the ultimate precision on the neutrino directions it is important to follow such movements.
An acoustic positioning system \cite{acoustic} and compasses within the electronics container provide 
a few centimetre and few degree precision on the Optical Module position and orientation.

The data communication and the power distribution to the lines are provided via a network of interlink cables and a 
junction box on the sea floor. Power is transmitted in alternating current from shore, via a single conductor in a 42 km 
long electro-optic cable, to a transformer in the junction box; the power return is via the sea. 

The data acquisition system is based on the `all-data-to-shore'' concept. In this mode, all signals from the PMTs 
that pass a preset threshold (typically 0.3 single photo electrons) are digitised in a custom built ASIC chip
and all digital data are sent to shore, via an optical Dense Wavelength Division 
Multiplexing system. On shore the data are processed in real-time by a farm of commodity PCs. 
The data flow is typically a couple of Gb/s and is dominated by the light generated by K40 decays of the salt in the seawater 
(typically a 50 kHz baseline singles rate per photo-multiplier).

\section{Event Reconstruction}

The main channel for the search for astrophysical point-like sources of neutrinos is the muon neutrino. 
The high rate of downgoing $\mu$  from the interactions of cosmic rays in the atmosphere restricts 
such searches to events coming from below, or only a few degrees above, the horizon. 
The remaining background is then the flux of atmospheric $\nu_{\mu}$ and those few remaining atmospheric 
$\mu$ events misreconstructed as upgoing. 
The long scattering length of blue light in seawater provides an excellent directional resolution on 
the $\nu_{\mu}$ primary of 0.38$^\circ$ for an $E^{-2}$ source \cite{angres}.
Thus, angular clustering requirements yield a strong suppression of both backgrounds. 

The effective area of neutrino telescopes such as ANTARES and IceCube to cascade events
(neutral-current (NC) interactions, and $\nu_e$ and $\nu_{\tau}$ charged-current (CC) interactions) 
is generally lower than that of $\nu_\mu$ CC interactions, due to the shorter range of the outgoing lepton. 
Additionally, the angular resolution of the cascade channel is inferior. Nevertheless, it has 
several advantages: neutrino events are more easily distinguished from the background of atmospheric
muons, allowing both up- and down-going events to be studied. Furthermore, the energy deposited in the
detector is better correlated with the energy of the primary neutrino. 

\begin{figure}[t]
\centerline{
  \minifigure[Angular resolution for cascade topology events.]
     {\includegraphics[width=5.5cm]{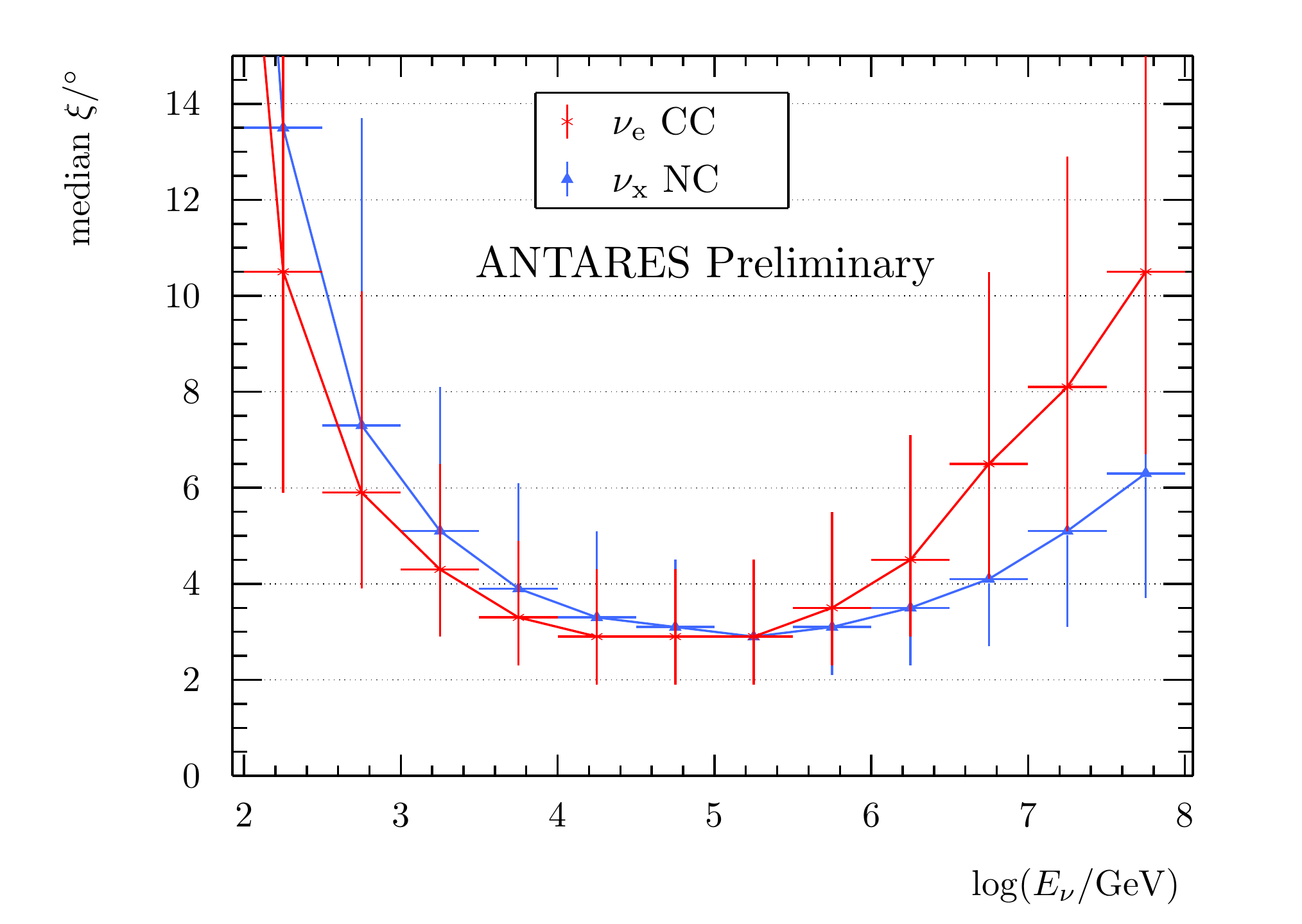}\label{cascade-ang}}   
  \hspace*{4pt}
 \minifigure[Energy resolution for cascade topology events.]
     {\includegraphics[width=5.5cm]{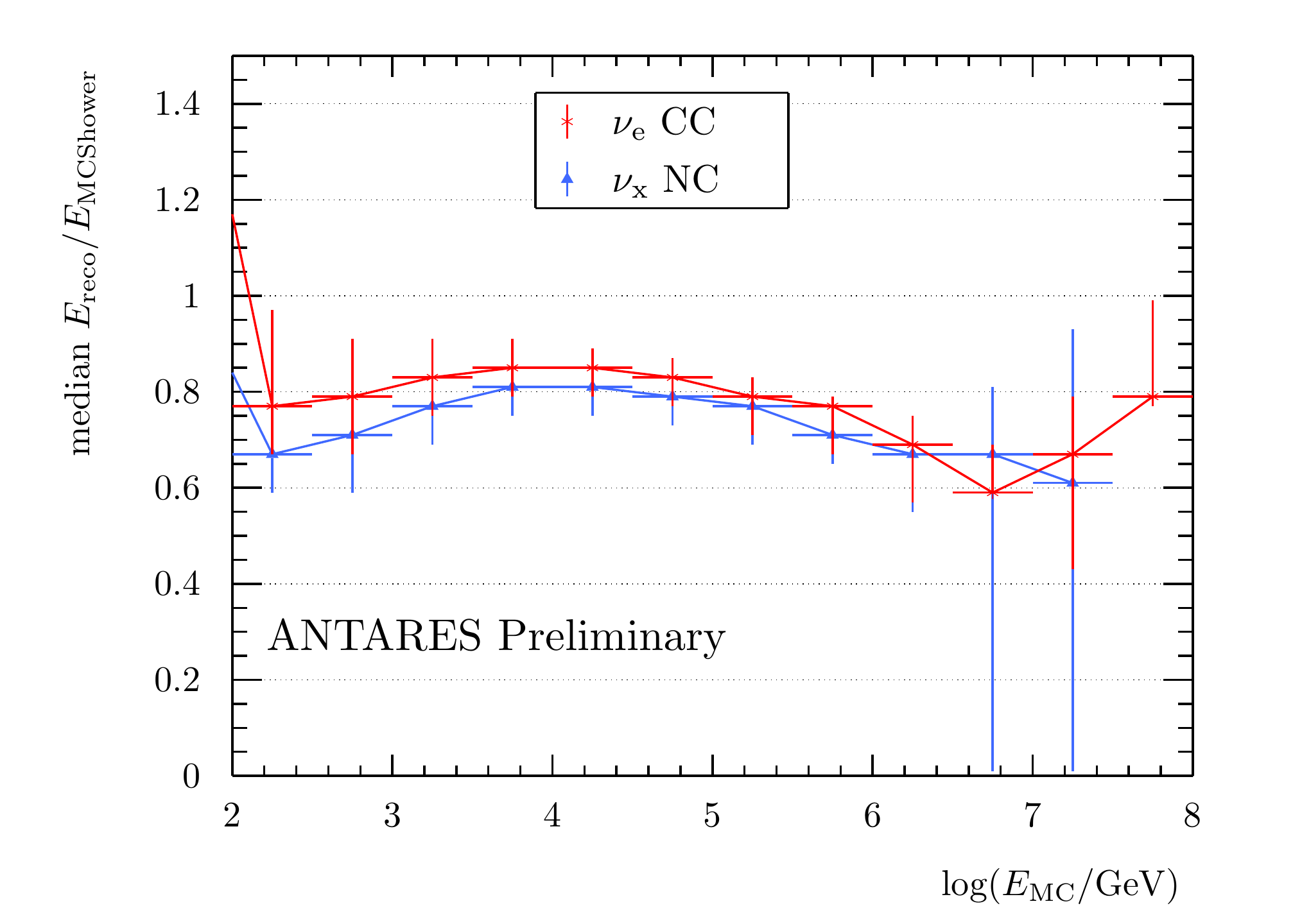}\label{cascade-energy}}
}
\end{figure}

The performance of the current ANTARES cascade reconstruction algorithm \cite{ICRC1078} 
yields a median angular resolution of typically $3^{\circ}$ for the energy range 10-100 TeV (\fref{cascade-ang}). 
The corresponding energy resolution is about 5\% (\fref{cascade-energy}) and thus is limited 
by the current systematic uncertainty of 10\%. 
Below 10 TeV, the resolution degrades due the reduced number of detected photons, 
while above 300 TeV the events start to saturate the detector.   

\section{Point source searches for astrophysical neutrinos}

\begin{figure}[t]
\centerline{
  \minifigure[Arrival directions of events in the all-sky point source analysis.]
     {\includegraphics[width=5.5cm]{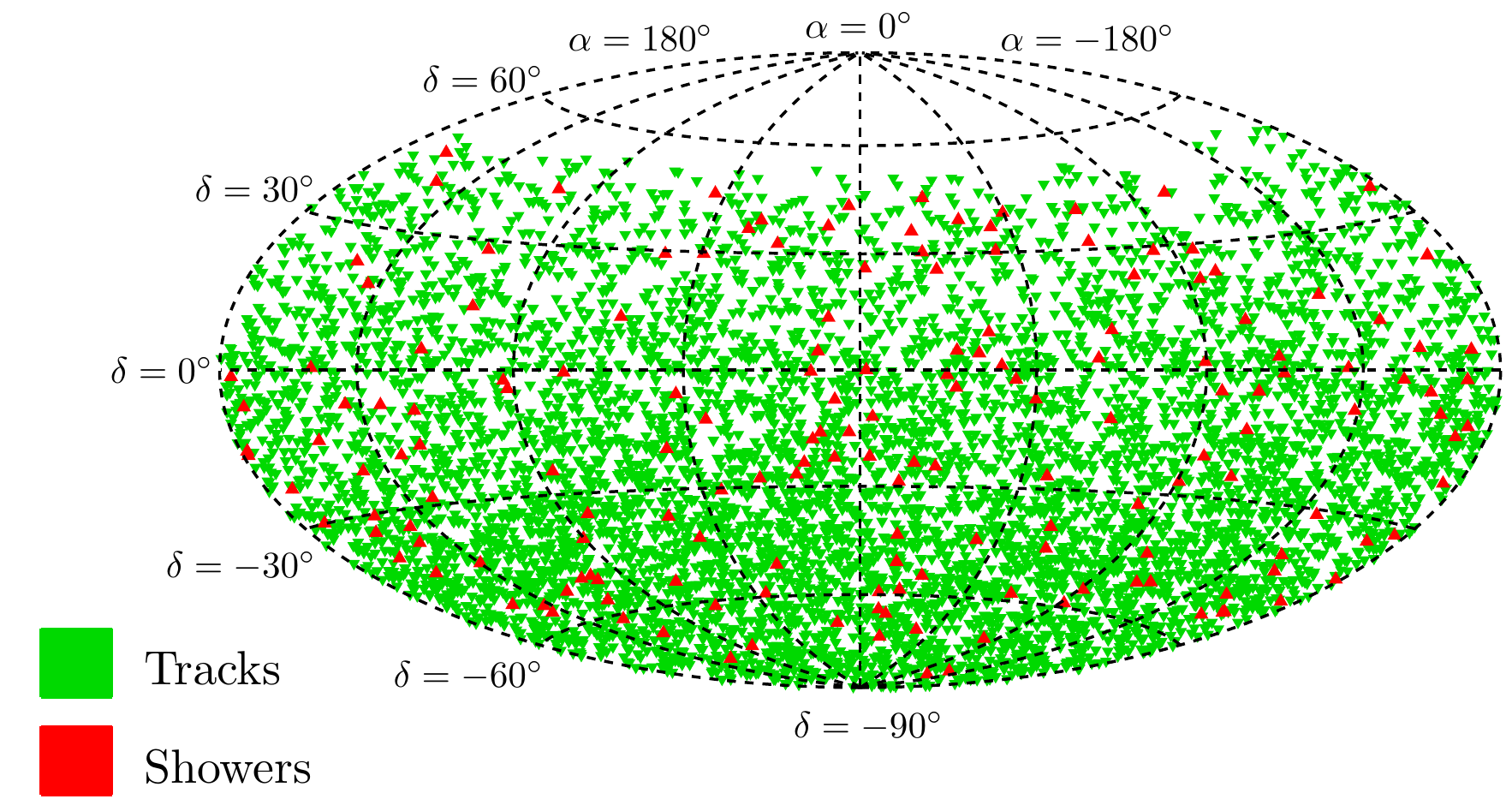}\label{ptsource-sky}}
  \hspace*{4pt}
  \minifigure[Limits and sensitivity of the ANTARES targeted source for flavour uniform 
  neutrino points sources with an $E^{-2}$ spectra in terms of flux per flavour.]
%     {\includegraphics[width=5.5cm]{ptsource-limit.pdf}\label{ptsource-limit}}
     {\includegraphics[width=5.5cm]{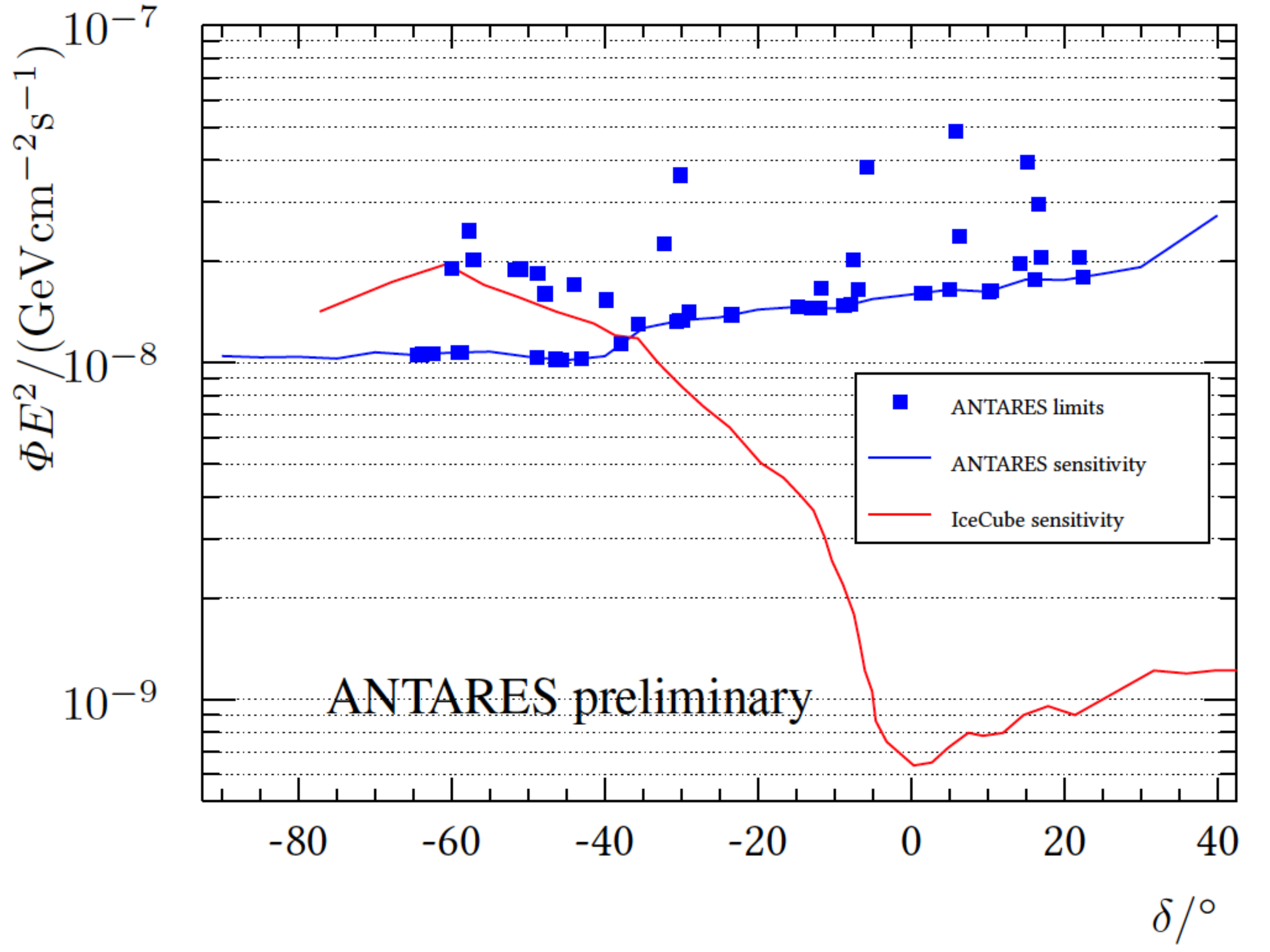}\label{ptsource-limit}}
}
\end{figure}

The inclusion of the cascade channel has allowed for the first time a combined point-source search using both 
muon-track and cascade events using 1690 days of effective livetime from 2007 to 2013 \cite{ICRC1078}.
After cuts, the sample consisted of 6261 muon-track events, and 156 cascade events, with an estimated
contamination of 10\% mis-reconstructed atmospheric muons in each. 
The resulting skymap is shown in \fref{ptsource-sky}. 

While the atmospheric background produces 
predominantly muon-track events, an $E^{-2}$ point source with a flavour-uniform flux would be expected to
produce a cascade-to-track ratio of 3:10, significantly increasing the sensitivity of the search.
The achieved search sensitivity was approximately $10^{-8} {\rm GeV}^{-1} {\rm cm}^{-2} {\rm s}^{-1}$  for $\delta < -40^\circ$
as shown in \fref{ptsource-limit}. An untargeted point-source search, a search over a list of pre-specified candidates, 
and a search using the origins of the IceCube events reported in Ref. \cite{ICflux} were applied to this data. 
No significant excess was observed. 

\subsection{Joint ANTARES/IceCube Point source search}

\begin{figure}[t]
\centerline{
  \minifigure[Fractional contribution of each data set to the total number of signal events passing cuts 
  in the joint ANTARES-IceCube analysis for an $E^{-2.5}$ spectra with no cutoff as a function of $\delta$.]
     {\includegraphics[width=5.5cm]{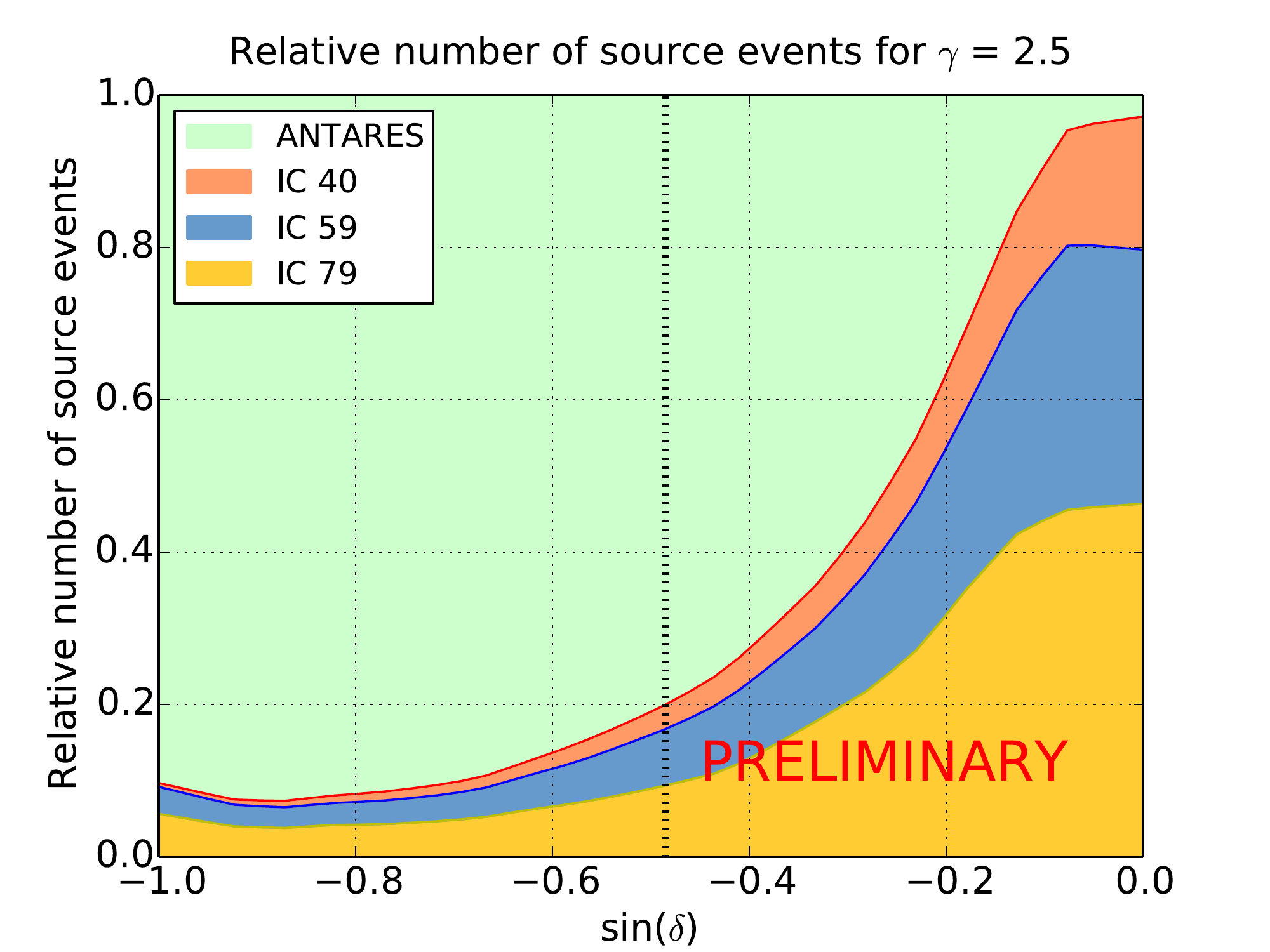}\label{ANTandIC-combined}}
  \hspace*{4pt}
  \minifigure[Sensitivity (lines) and limits (dots) of the joint ANTARES-IceCube analysis for 
  an $E^{-2.5}$ spectra with no cutoff as a function of $\delta$. ANTARES (blue), IceCube (red), combined (green).]
     {\includegraphics[width=5.5cm]{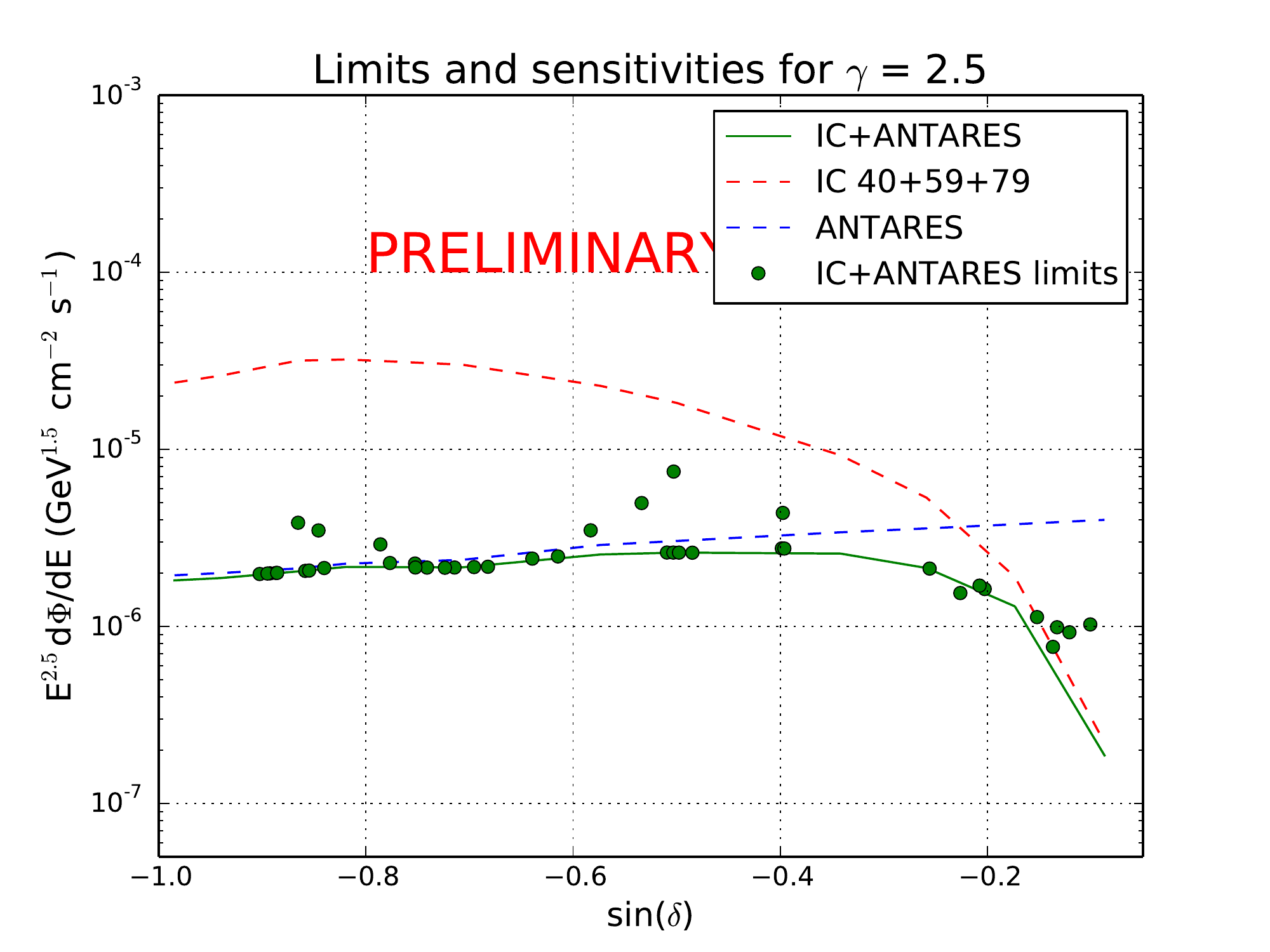}\label{ANTandIC-limit}}
}
\end{figure}

A joint analysis using ANTARES and IceCube data is detailed in Ref. \cite{ICRC1076}.
The fractional number of source events expected to be present in each data
set is shown in \fref{ANTandIC-combined} for the current best-fit to the IceCube flux. 
The results of the combined search are shown in \fref{ANTandIC-limit}, 
for an $E^{-2.5}$ source spectrum. The ANTARES contribution is dominant for declination $< -15^\circ$.
No significant cluster is found, with the most significant source on the candidate list being the blazar 3C 279,
with a pre-trial p-value of 5\%. The combined analysis improves on
the results from both experiments, indicating the complementarity of the two instruments.

\subsection{Limits on a point source origin of the IceCube signal}

\begin{figure}[hb]
\centerline{\includegraphics[width=5.5cm]{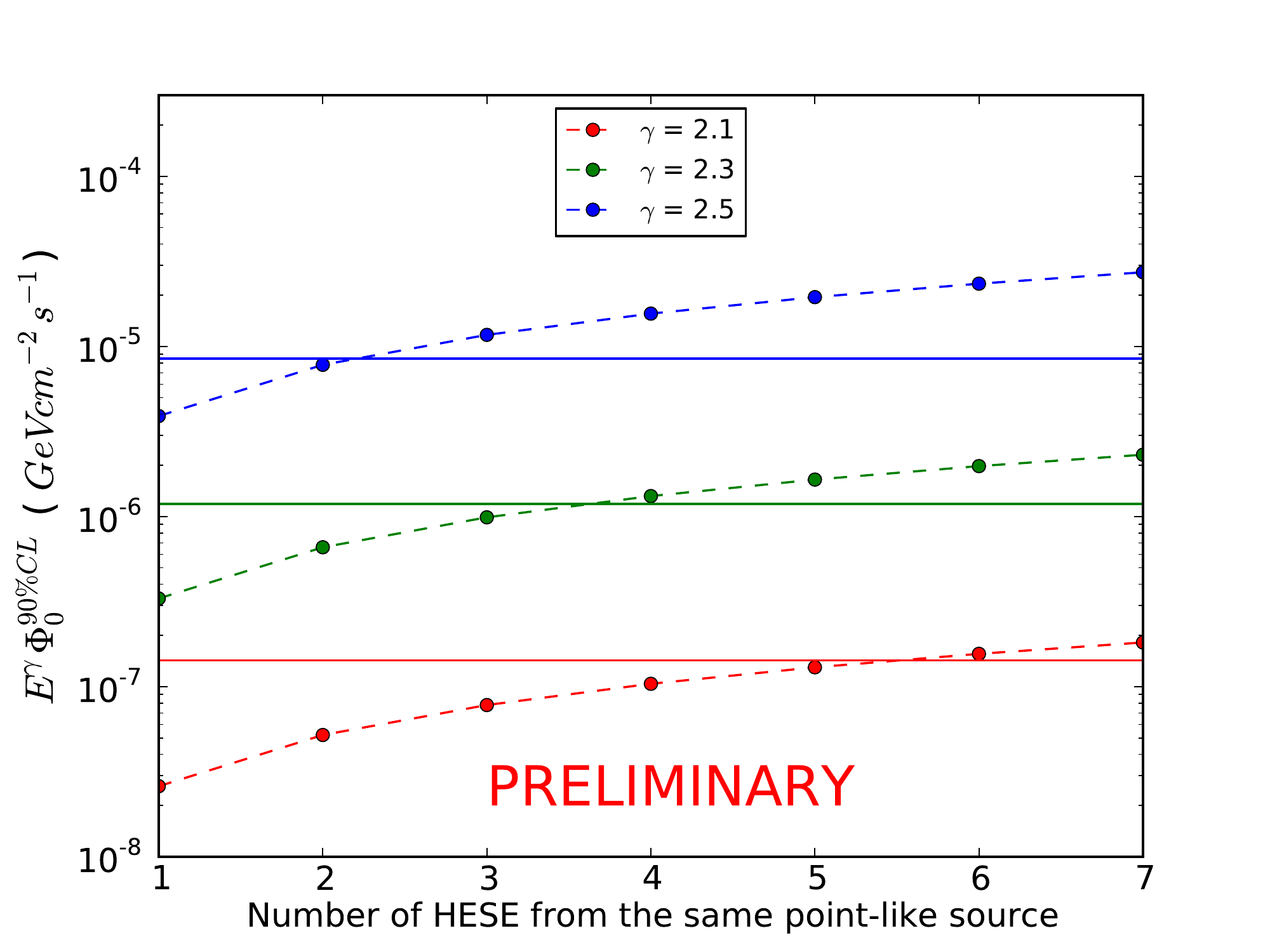}}
\caption{ANTARES limits (solid lines) at 90\% C.L. on the contribution of point-like sources to the IceCube 
HESE sample for various spectral indices, shown for $\delta=-29^\circ $. These are compared with the flux 
required to produce a given number of HESE events. Similar results are obtained for other declinations around
the Galactic Centre.}
\label{HESE-limit}
\end{figure}

It has been proposed \cite{Halzen} that the cluster of IceCube events seen in Ref. \cite{ICflux} 
could be due to a single point-like source, which is not detectable due to the poor angular resolution. 
The non-detection of an ANTARES point-like source in this region, as reported by Ref. \cite{ICRC1077},
limits the flux of such a source as a function of spectral index, shown by the solid lines and y-axis of
\fref{HESE-limit}. The flux required to produce a given number of events in the HESE analysis (x-axis) is also
shown. The range where the latter is greater than the former rules out a corresponding contribution
from any single point-like source with that spectral index at 90\% C.L..
The result above is particularly relevant because the current best-fit spectrum (between 25 TeV
and 2.8 PeV) of the IceCube flux has a spectral index of $-2.5 \pm 0.09$  \cite{ICflux2}.
ANTARES can thus rule out any single point-source of neutrinos in the region of the Galactic Centre with 
spectral index of -2.5 as having a flux corresponding to more than 2 HESE events.

\section{Transient sources}

\subsection{Flares from AGN and X-ray binaries}

\begin{figure}[hb]
\centerline{\includegraphics[width=10cm] {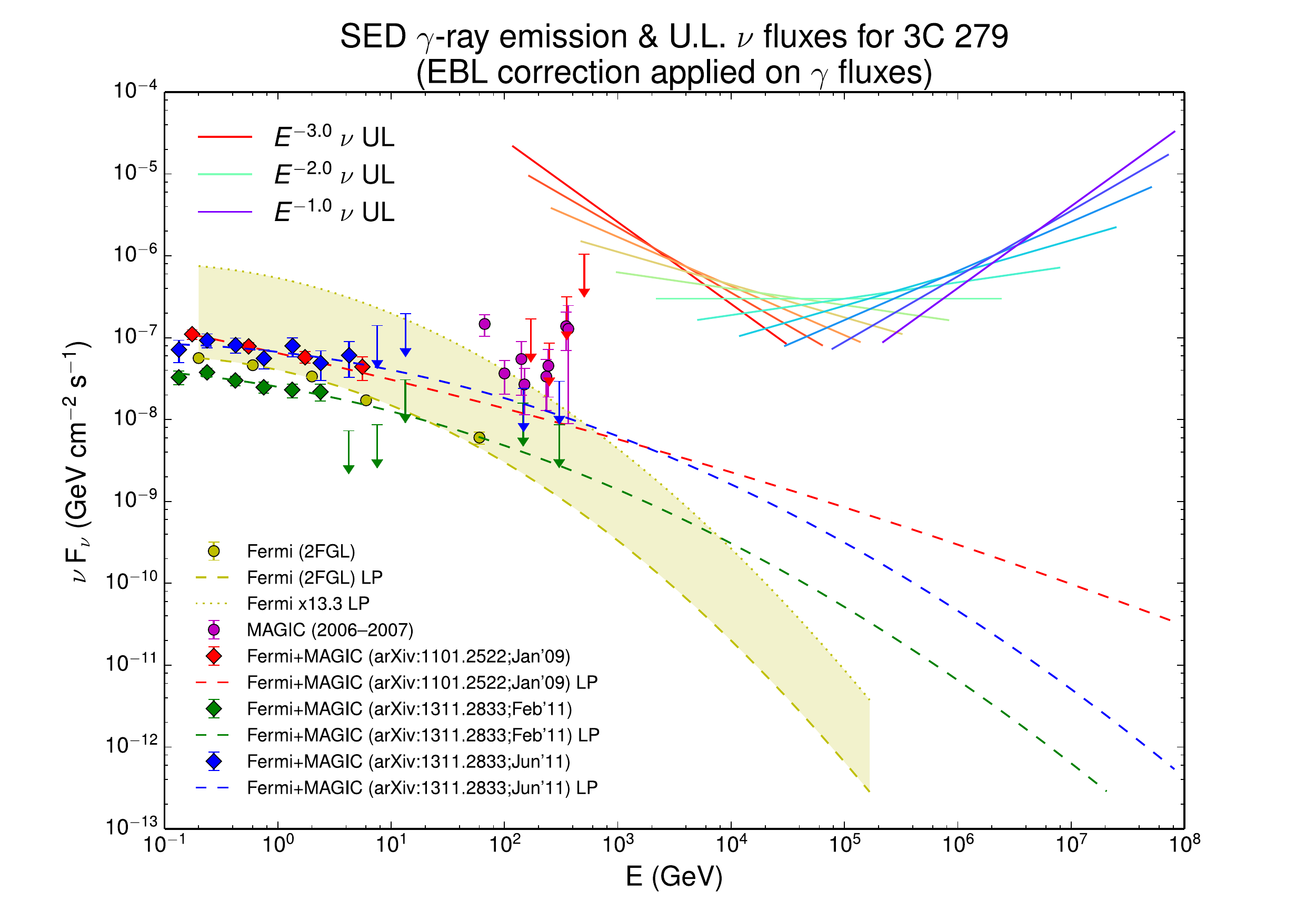} }
\caption{Limits on the neutrino flux from the blazar 3C279 as a function of spectral index (solid lines), compared
 to the observed (points) and extrapolated (dashed lines) gamma-ray spectra observed by FERMI and IACTs.}
\label{3C279-limit}
\end{figure}

Active galactic nuclei (AGN) have long been proposed as a source of high-energy cosmic rays and, hence, neutrinos \cite{Gaisser}.
Blazars, AGNs with jets pointing towards the line-of-sight, exhibit bright flares which dominate the 
extragalactic gamma-ray sky observed by Fermi-LAT \cite{Fermi}.

Using multi-wavelength observations, several bright blazars have been reported 
by the TANAMI Collaboration \cite{Tanami}
to lie within the 50\% error bounds of the reconstructed arrival directions of the PeV-scale events IC 14 and 
IC 20 observed by IceCube. ANTARES observes signal-like 
events from the two brightest blazars, both in the field of IC 20 \cite {Kadler}, although this is also consistent with 
background fluctuations. A lack of such events from the field of IC 14 excludes a neutrino spectrum softer than 
$E^{-2.4}$ as being responsible for this event. The highest-energy event IC 35 (``Big Bird'') was detected during 
an extremely bright flare from the blazar PKS B1424-418, which lies within the 50\% error region of the IC 35 
arrival direction. ANTARES finds only one event within $5^\circ$ of this source during the flaring period, 
whereas approximately three would be expected from random background fluctuations alone.

In another analysis \cite{ICRC1075}, ANTARES targets a sample of 41 blazar flares 
observed by Fermi LAT and 7 by the IACTs H.E.S.S., MAGIC, and VERITAS. The lowest pre-trial p-value of 3.3\% 
was found for the blazar 3C 279, which comes from the coincidence of one event with a 2008 flare previously 
reported in Ref. \cite{ANTblazar}. However, the post- trial p-value is not significant. The resulting limits are given in 
\fref{3C279-limit}.

Similar methods were also used to search for neutrino emission during the flares from 
galactic x-ray binaries \cite{ANTbinaries}. A total of 34 x-ray- and gamma-ray-selected binaries were studied, 
with no significant detections, allowing some of the more optimistic models for hadronic acceleration in these 
sources to be rejected at 90\% C.L..

\subsection{Gamma ray bursts}

\begin{figure}[t]
\centerline{
  \minifigure[
  Range of jet $\Gamma$-factor and baryonic loading fp excluded by ANTARES 
in the case of GRB110918A using the NeuCosmA model of Ref. \cite{NeuCosmA},
as described in Ref. \cite{ICRC1057}.  
The assumed values of $\Gamma = 316$ and $fp = 10$ are shown by the red point, 
 while the colour-coding gives the expected number of observable neutrinos.]
  {\includegraphics[width=5.5cm,clip=true, trim=0.5cm 8cm 2cm 8.5cm] {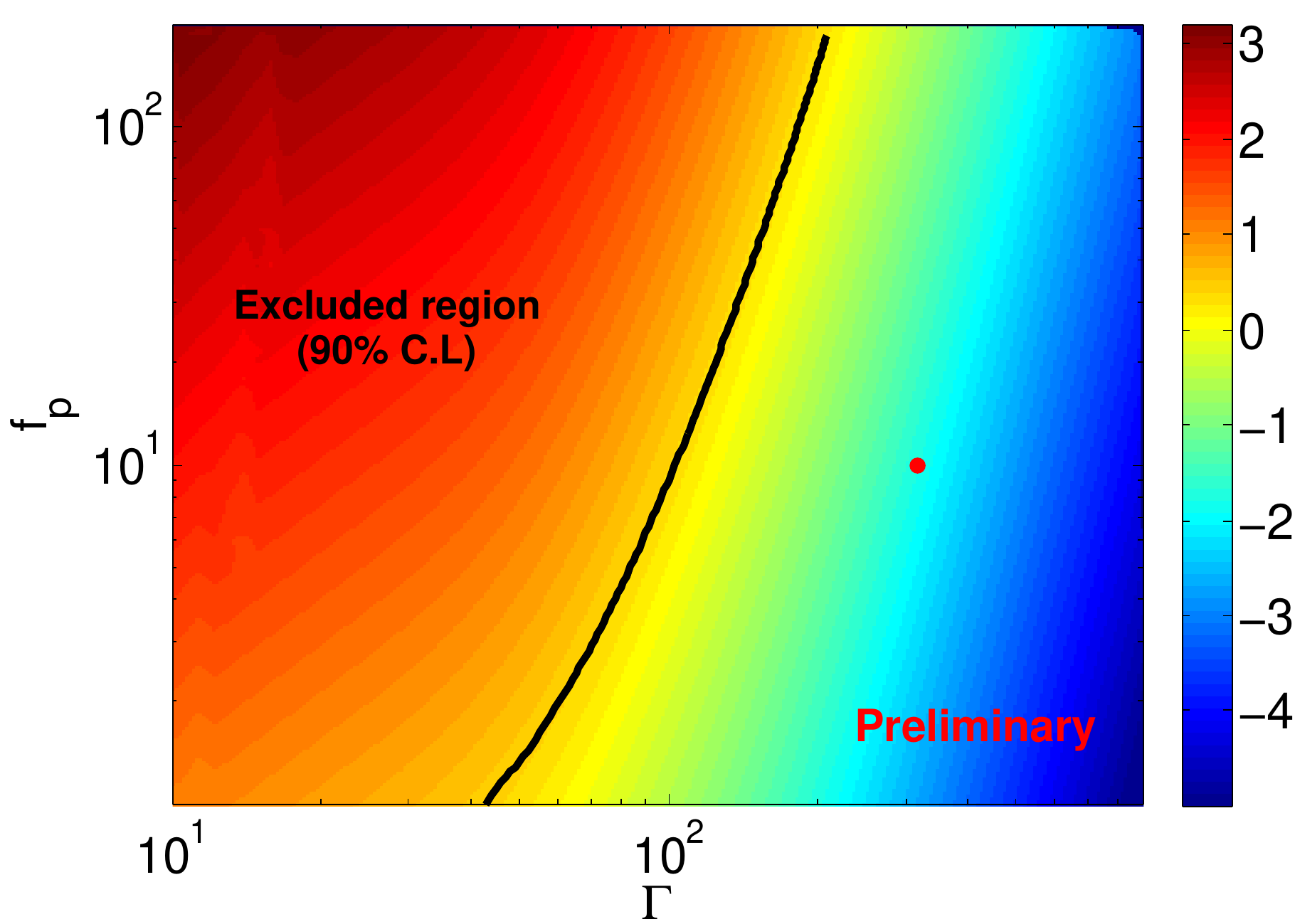} \label{GRB110918A}}
  \hspace*{4pt}
  \minifigure[Limiting magnitudes and delay times of optical follow-up observations to ANTARES alerts 
  with ROTSE and TAROT \cite{ICRC1093} compared to (grey lines) the light-curves from measured GRBs.]
  {\includegraphics[width=5.5cm] {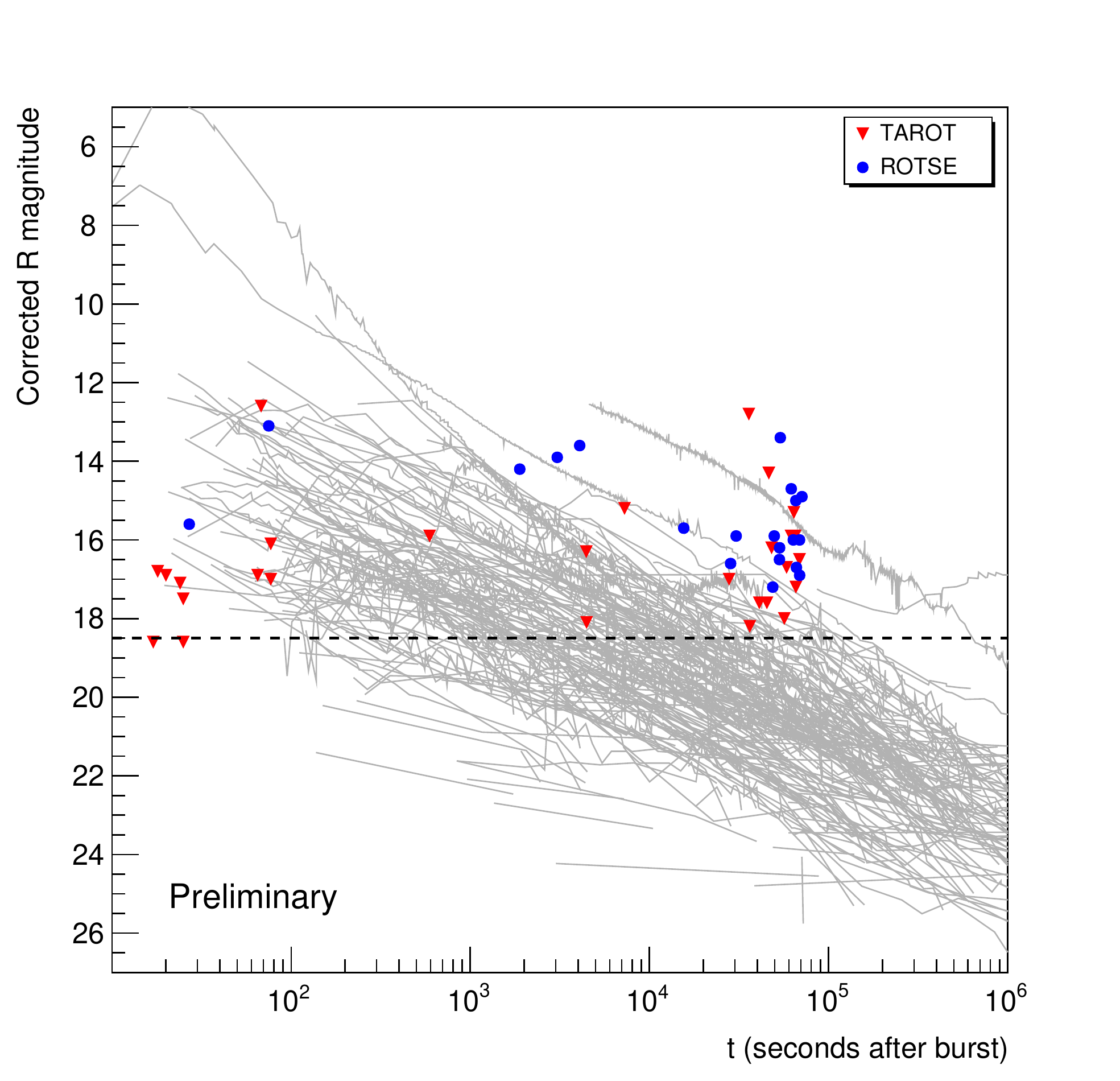} \label{tatoo}}
}
\end{figure}

Long-duration gamma-ray bursts (GRBs) have been proposed as a 
source of the highest-energy cosmic rays \cite{Waxman}. 
ANTARES has searched for a neutrino flux from GRBs considering two models of the emission processes: 
the NeuCosmA description of Ref. \cite{NeuCosmA} and the ``photospheric'' model of Ref. \cite{photospheric}.
In each case, the expected signal is simulated on a burst-by-burst basis, and the detector response and background 
are modelled using the exact detector conditions at the time of the burst. The ANTARES analysis using the 
NeuCosmA model was developed and applied to a sample of 296 bursts in Ref. \cite{ANTgrb}, with no coincident 
neutrino  events detected. 
Since then, one especially powerful burst GRB110918A, and the nearby burst GRB130427A, have 
been identified as promising candidates for neutrino detection, and studied in detail in Ref. \cite{ICRC1057}.
No coincident events are observed from either burst. 
The predicted $\nu$ emission scales with $\Gamma^{-5}$ and linearly with $f_p$, 
allowing limits to be set on the bulk gamma-factor and baryonic loading of the jet, as shown in \fref{GRB110918A}. 

A search using the photospheric models is developed in Ref. \cite{ICRC1068}, and will shortly be 
unblinded. The GRB search methods are also being extended to test Lorentz invariance violation 
\cite{ICRC1057}, which would delay the arrival times of TeV neutrinos compared to GeV photons.

\subsection{Optical and X-ray follow-up}

The TAToO (telescopes-ANTARES target-of-opportunity) program \cite{tatoo} performs 
near-real- time reconstruction of muon-track events. 
If a sufficiently high energy event is reconstructed as coming from below the horizon (i.e. those 
events most likely to be of astrophysical origin), an alert message is generated to trigger robotic optical telescopes, 
and, with a higher threshold, the Swift-XRT. The very short alert-generation time (a few seconds) and half-sky 
simultaneous coverage of ANTARES makes it ideal for detecting transient signals, and optical and x-ray follow up 
observations have been initiated within 20 s and one hour respectively.

Results from 42 optical and 7 x-ray alerts have been analysed. While no associated transient 
event was detected, this non-observation can be used to place limits on the astrophysical origin of the detected 
neutrinos \cite{ICRC1093}, as shown in \fref{tatoo}. The steep fall-off of the light-curves emphasises the 
need for a rapid alert generation and follow-up: observations within one minute can rule out a GRB origin with high 
confidence, while those after one day would be unlikely to detect even a bright GRB.

\section{Diffuse flux searches}

\begin{figure}[hb]
\centerline{\includegraphics[width=8cm] {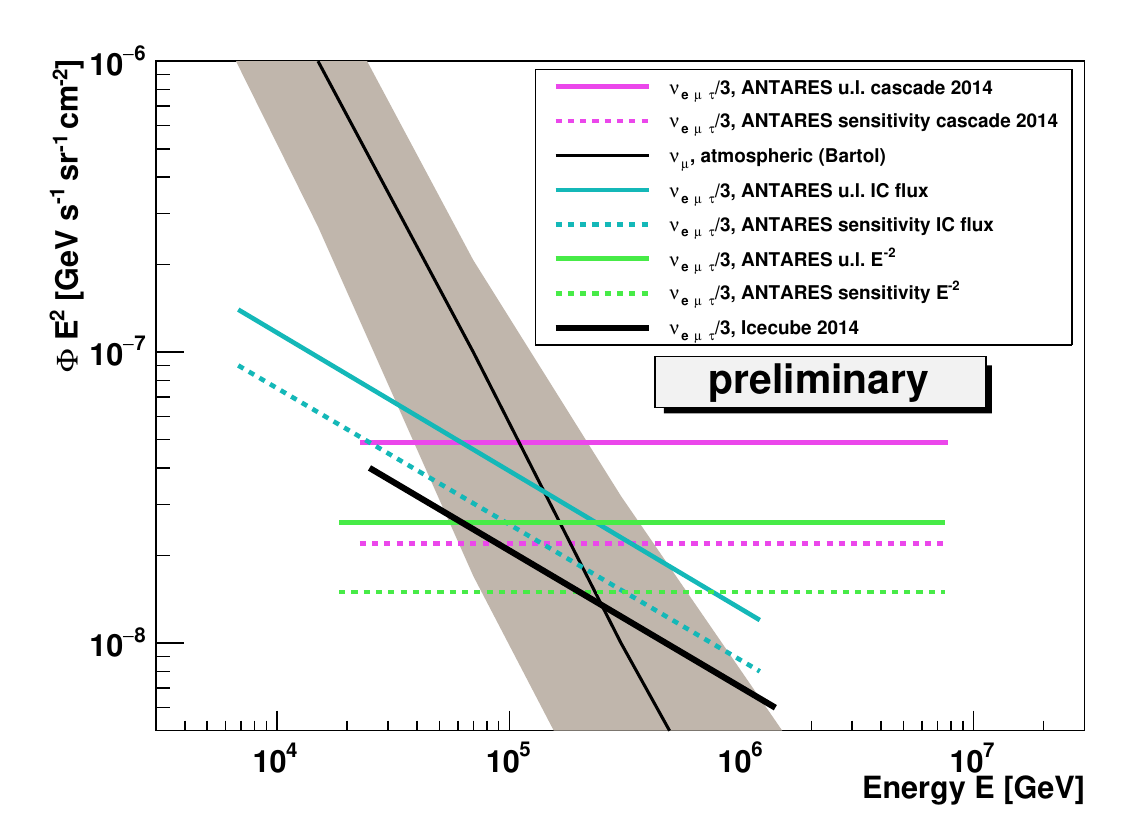} }
\caption{ANTARES sensitivity to (dotted), and limits on (solid), the diffuse astrophysical neutrino
measured by IceCube \cite{ICRC1065}.  Shown are (pink) the previous ANTARES limit on an $E^{-2}$ spectrum \cite{ANTdiffuse},
and current results on (blue) the flux (thick black line) observed by IceCube \cite{ICflux2} and (green) an $E^{-2}$ spectrum. 
This is compared to the conventional atmospheric background flux (thin black line) \cite{Honda}, 
with associated error (grey shading).}
\label{diffuse-limit}
\end{figure}

A diffuse flux search in ANTARES has been developed that makes optimal uses of both muon track and cascade 
events \cite{ICRC1065}. Since any explicit selection of muon-like and cascade-like events 
inevitably discards events with topologies falling between the two classes, no such selection was made. The 
procedure was applied to 913 days of effective livetime between 2007 and 2013. 
The expected number of events from the 
standard and prompt atmospheric background \cite{Honda, Enberg} 
was $9.5\pm2.5$, composed of 5.5 $\nu_\mu$ CC, 1 atmospheric $\mu$, and 2.9 $\nu$ NC and $\nu_e$ events. 
The expectation from the IceCube neutrino flux reported by  \cite {ICflux} was $5.0\pm1.1$ 
events. After unblinding, 12 events passed the selection cuts--consistent with both background only, and background 
and IceCube diffuse flux expectations. The resulting limits on an $E^{-2}$ flux are given in \fref{diffuse-limit}.

\section{Extended source searches}
In addition to the numerous point-like candidate neutrino sources, several extended regions 
have been proposed as hadronic acceleration sites. ANTARES searches for an excess neutrino flux from these 
regions using ``on-zones'' defined by specific templates, which are compared to ``off-zones'' of exactly the same size 
and shape, but offset in right ascension. Thus the off-source regions give an unbiased estimate of the background in 
the source region in a way that is independent of simulations. Results for the Fermi Bubbles, Galactic plane, and the 
IceCube cluster are described below.

\subsection{Fermi bubbles}

\begin{figure}[t]
\centerline{
  \minifigure[ On- and off-zone search regions for the Fermi Bubble search of Ref. \cite{ICRC1059}, 
  compared to the ANTARES visibility (blue shading).]
  {\includegraphics[width=5.5cm] {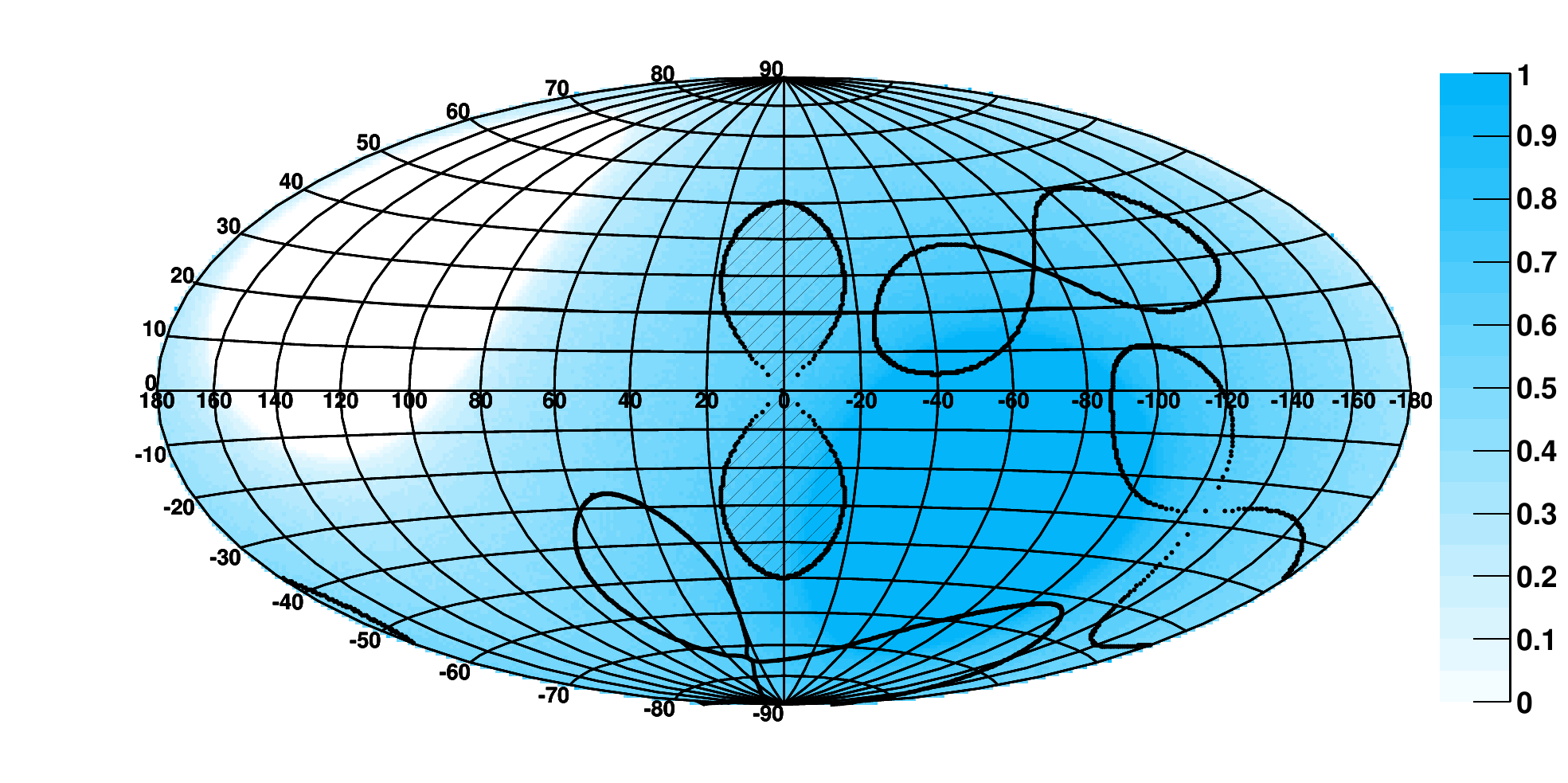} \label{FB-sky}}
  \hspace*{4pt}
  \minifigure[90\% C.L. upper limits (lines) on the neutrino flux from the Fermi Bubbles, 
  compared to (shaded regions) expectations \cite{Lunardini} for different spectral shapes.]
  {\includegraphics[width=5.5cm] {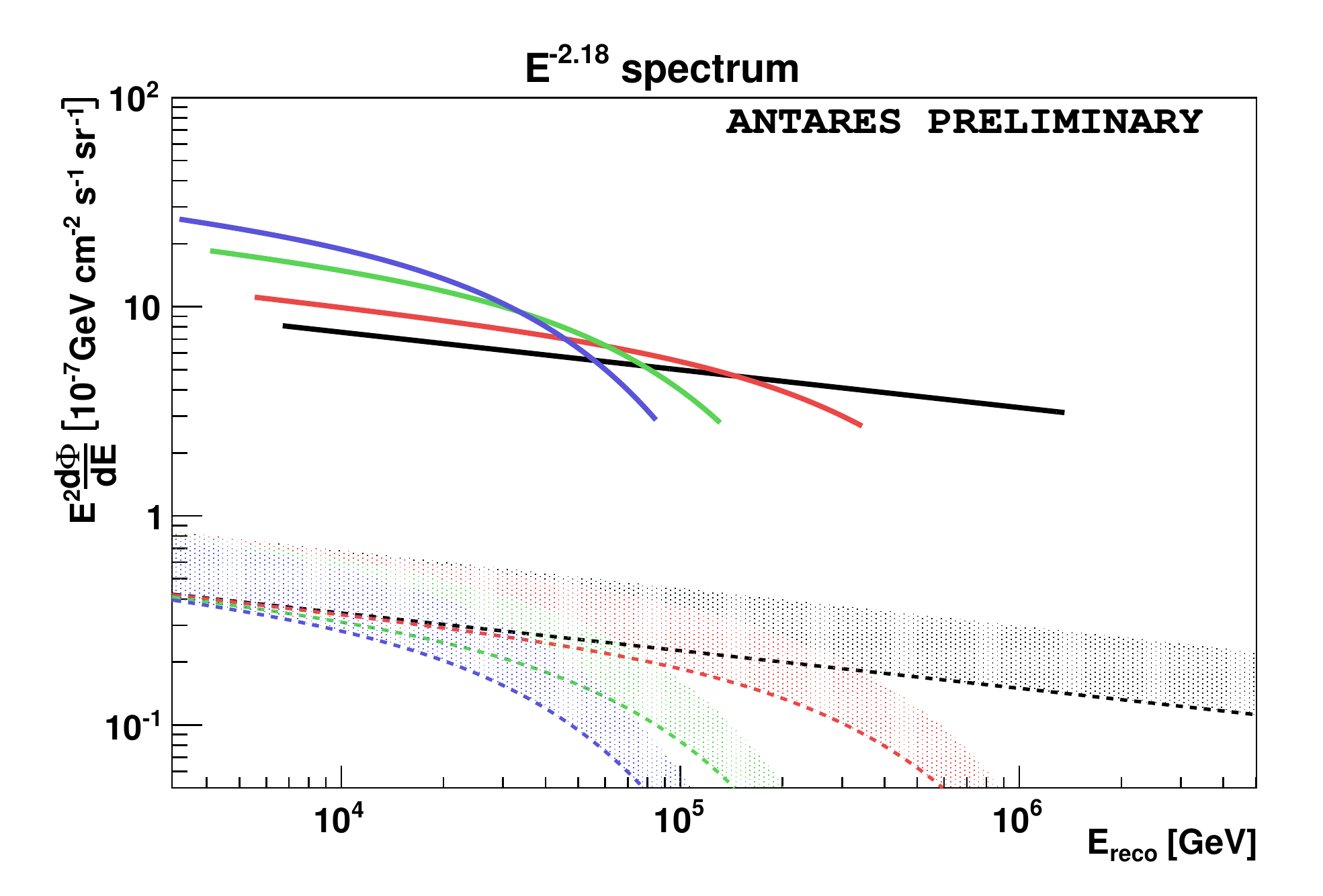} \label{FB-limit}}
}
\end{figure}

The Fermi Bubbles \cite{FBubbles} are large regions of $\gamma$-ray emission 
extending perpendicular to the Galactic Centre and have been proposed as sites of hadronic acceleration \cite{Lunardini}, with neutrinos expected from p-p collisions. 
A first search in ANTARES data from 2008-2011 for emission from these regions was presented in Ref. \cite{ANTfb} - here, 
an update is presented adding 2012-2013 data.

The on- and off-zone regions used in the Fermi Bubble analysis are shown in \fref{FB-sky}. Flavour-uniform $E^{-2}$ 
and $E^{-2.18}$ neutrino fluxes are assumed, where the latter is motivated by the best-fit proton spectrum 
of $E^{-2.25}$ reported by Ref. \cite{Lunardini}. 
Exponential cut-offs at energies of 500, 100, and 50 TeV are also tested.
A slight excess is found in the source region, corresponding to a 1.9 $\sigma$ significance. 
The corresponding upper limits on an $E^{-2.18}$ neutrino flux are compared in \fref{FB-limit} to the expectations from Ref. \cite{Lunardini}.

\subsection{Galactic plane}

\begin{figure}[hb]
\centerline{\includegraphics[width=6cm] {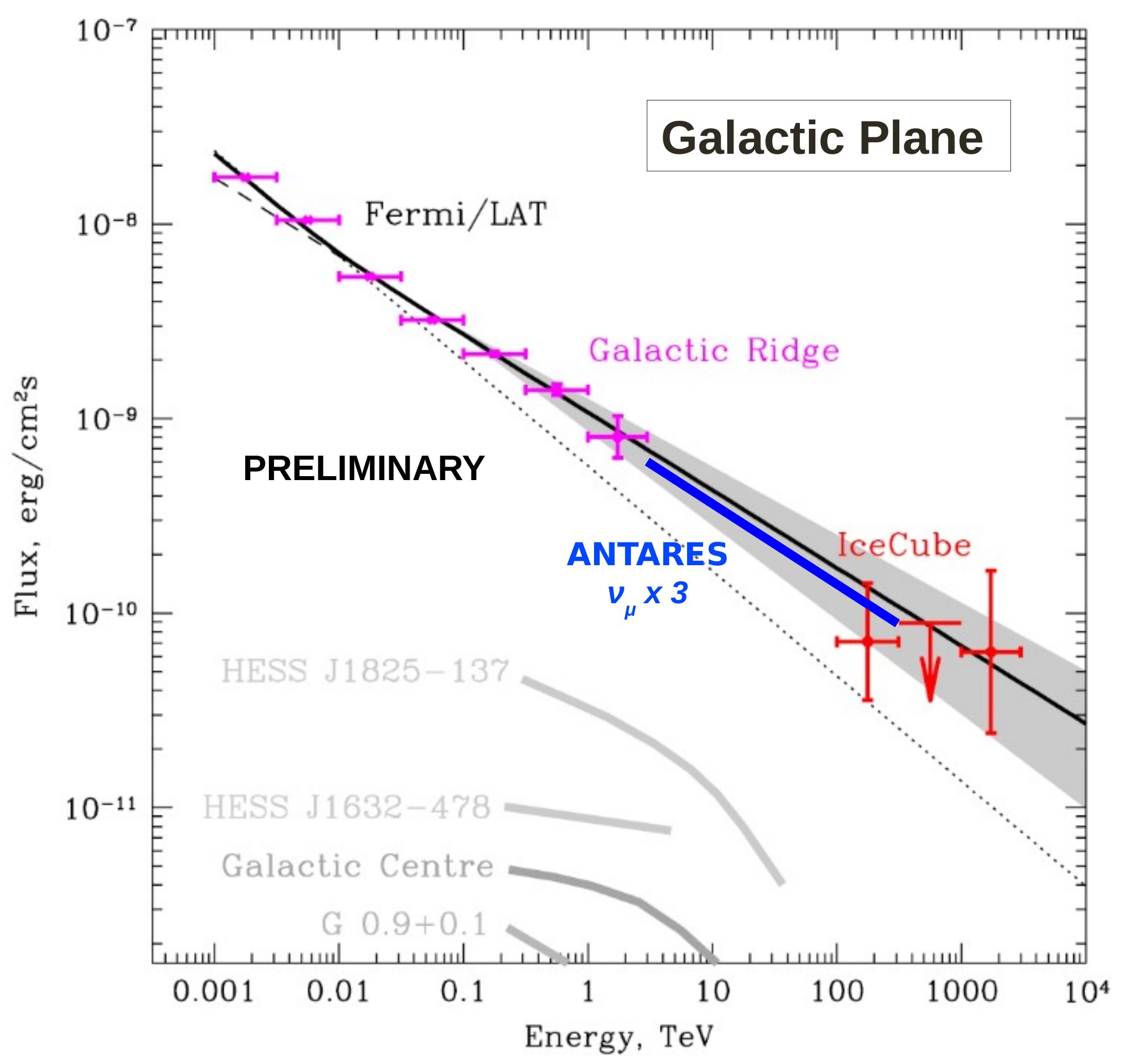} \includegraphics[width=6cm] {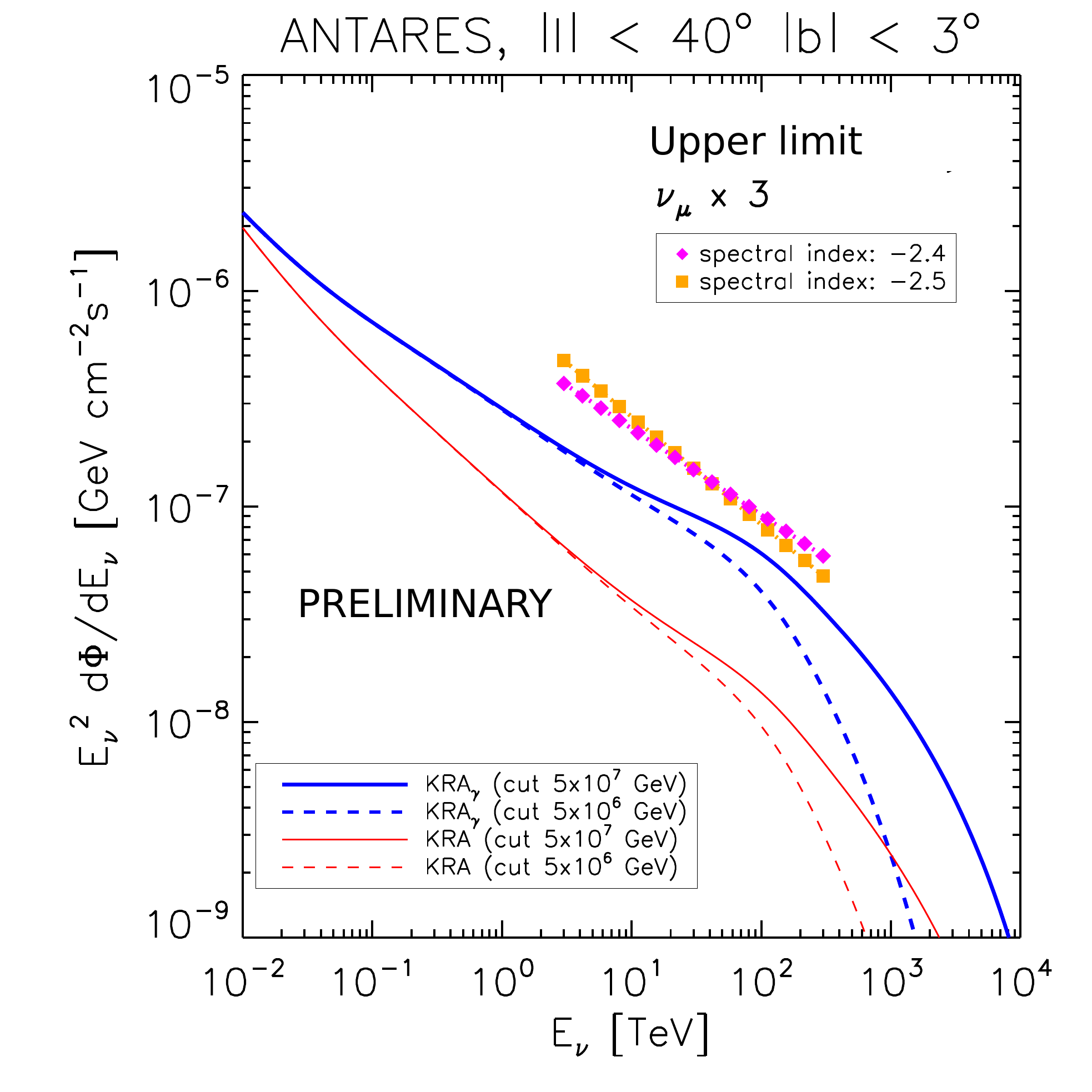}}
\caption{
ANTARES 90\% C.L. upper limits for the search for an excess of events from the central Galactic region \cite{ICRC1055}.
Left: for an $E^{-2.4}$ neutrino spectra, compared to the expected neutrino flux (solid black line)
extrapolated from the FERMI-LAT diffuse gamma-ray up to high energies. 
Right: for $E^{-2.4}$ and $E^{-2.5}$ neutrino spectra, compared to the predicted neutrino fluxes from Ref. \cite{Gaggero}.
}
\label{ridge-limit}
\end{figure}

Cosmic rays in our galaxy will collide with the interstellar medium to produce pions and, hence, neutrinos. 
Direct evidence for these processes comes from observations by Fermi-LAT \cite{Ackermann} of the diffuse galactic 
gamma-ray background. It is also interesting that the number of IceCube high energy starting events (HESE) 
in the $E > 100$ TeV range with angular directions consistent with this region 
corresponds to a flux consistent with that observed in $\gamma$-rays \cite{Neronov}, as shown in \fref{ridge-limit}. 
The large uncertainty in the arrival directions of cascade-like HESE, and their low number, 
makes this comparison difficult however.

The ANTARES northern latitude is ideally suited for studying the expected neutrino flux from the inner galactic plane. 
A search has been performed in the regions of galactic longitude $|l| < 40^\circ$ and 
latitude $|b| < 3^\circ$, as reported in \cite{ICRC1055}. 
The search used nine off-zones and one on-zone, and found no excess in the on-zone region (one event compared 
to an average of 2.5 for the off-zones). The resulting limits are shown in \fref{ridge-limit}. 
In particular, the hypothesis of a $1-1$ relation between the $\gamma$-ray and neutrino flux from the Galactic Ridge 
is ruled out at 90\% confidence, indicating that ANTARES is already testing the well-established 
multimessenger $\gamma$-$\nu$-CR paradigm in our galaxy. The limits cannot rule out however models from 
more-detailed simulations of galactic cosmic-ray propagation.

\subsection{IceCube cluster}

\begin{figure}[hb]
\centerline{\includegraphics[width=6cm] {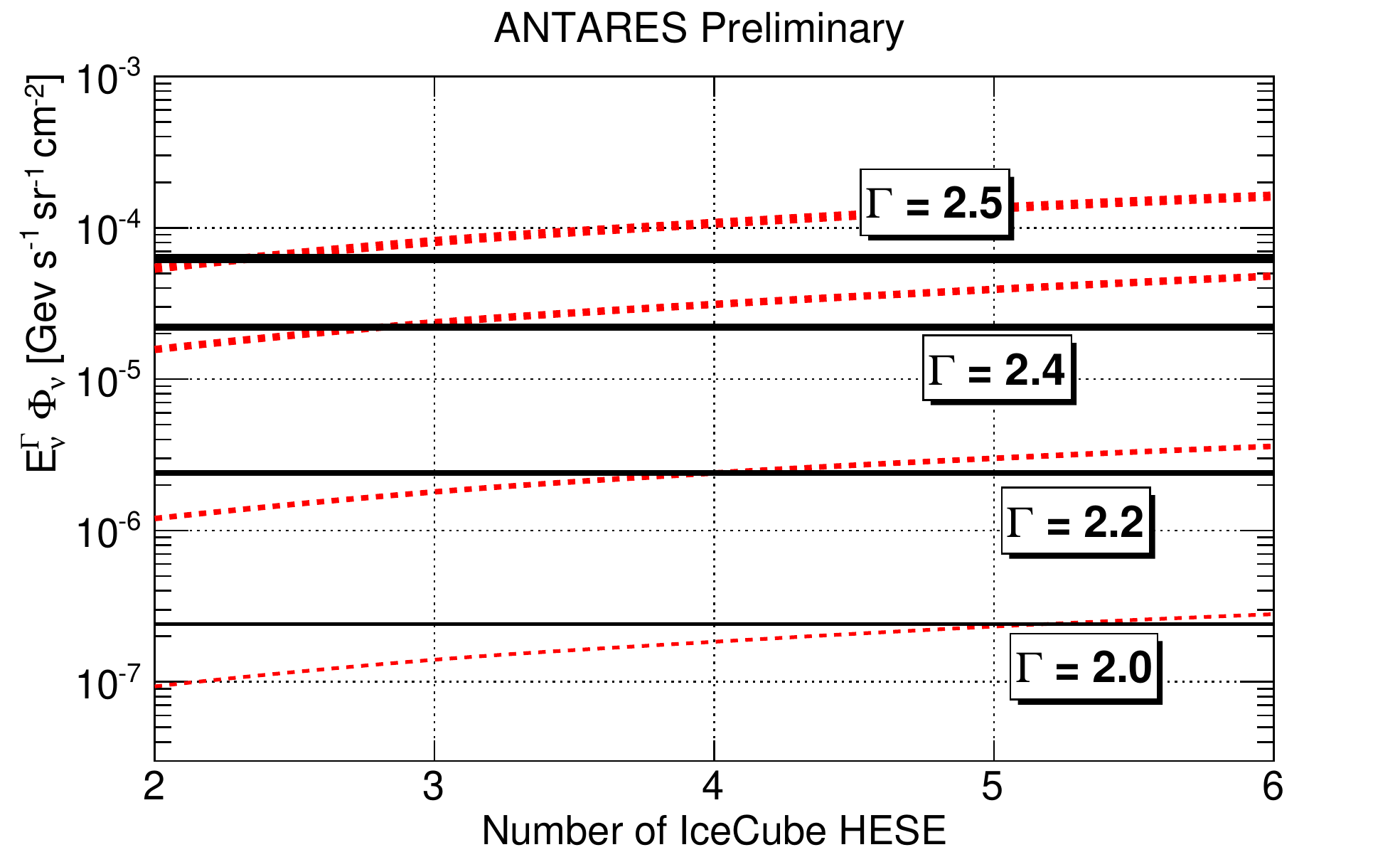} }
\caption{
ANTARES upper limits at 90\% C.L. (black) on a flavour-uniform neutrino flux from the
IceCube cluster region as a function of the spectral index $\Gamma$, compared to (red) the flux required to produce
an expected number of events in the IceCube HESE analysis \cite{Spurio}. The maximum number of IceCube
events allowed at 90\% C.L. is indicated by the crossing points of the red and black lines for a given spectral
index. See Ref. \cite{ICRC1055} for details.
}
\label{cluster-limit}
\end{figure}

The same techniques employed in the galactic plane search were used to probe the origin of the cluster of 
IceCube events reported in Ref. \cite{ICflux}. The analysis of Ref. \cite{ICRC1055} 
used twelve off-zones and one on-zone to search for an excess of events. 
One event passing the selection cuts is observed in both the on-zone and the 
average off-zone, i.e. no excess is observed. Resulting limits on the maximum number of HESE events 
produced by a  source with different spectral indices are presented in \fref{cluster-limit}. 
For the best-fit IceCube diffuse spectral index $\Gamma$ = 2.5 \cite{ICflux2}, 
ANTARES rejects at 
90\% confidence a flux from this region expected to produce three or more of the IceCube events in the cluster. This 
extends the results of Refs. \cite{angres, ICRC1077} for this region, which limit the existence of point-
like and mildly extended sources in this region.

\section{Dark matter and exotics}

\begin{figure}[t]
\centerline{
  \minifigure[ ANTARES limits on $\sigma^p_{SD}$ from the Sun as a function of the WIMP mass. ]
  {\includegraphics[width=5.5cm] {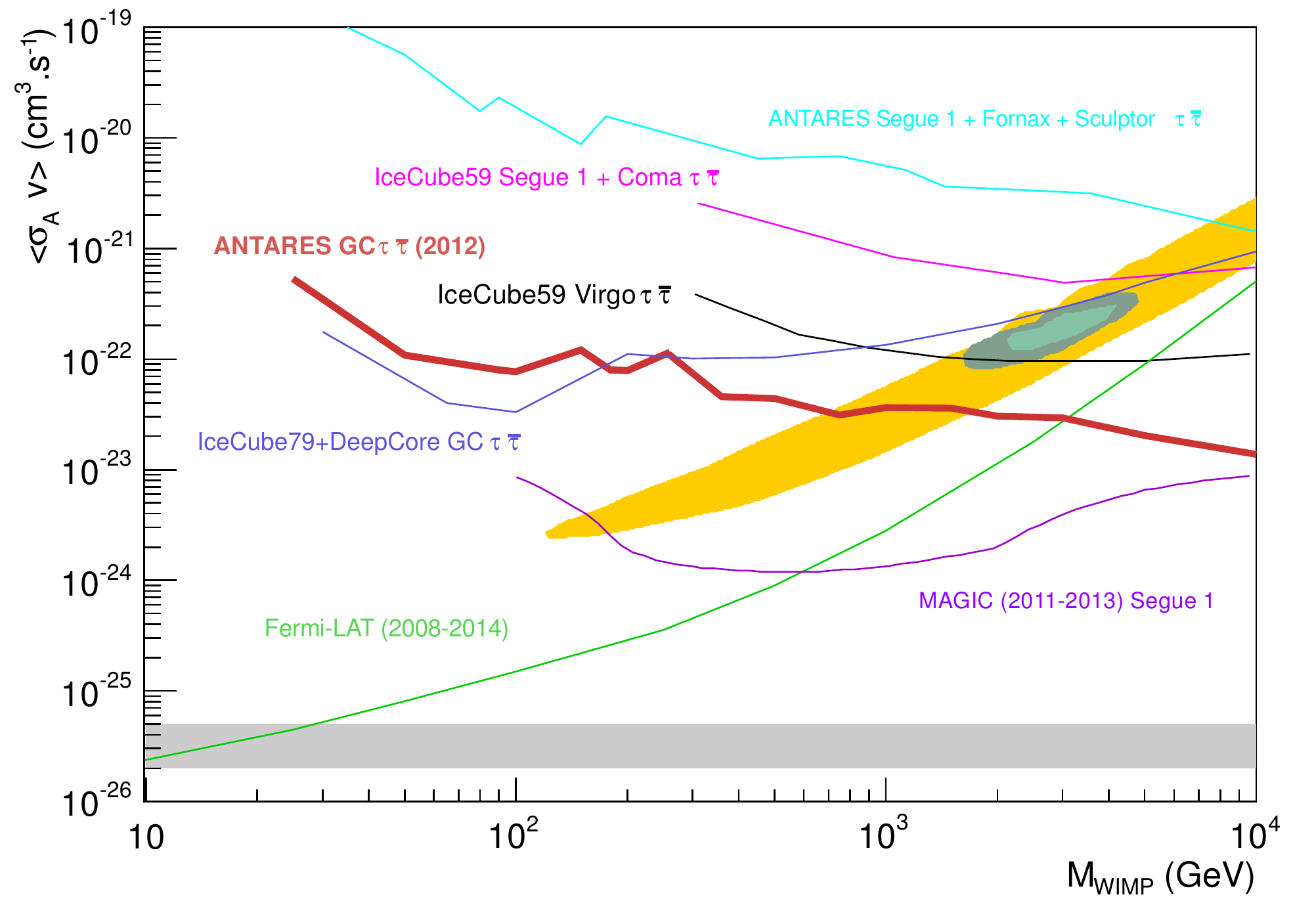} \label{dmsd} }
  \hspace*{4pt}
  \minifigure[ANTARES limits on $\sigma_A\nu$ from the Galactic Centre as a function of the WIMP mass. ]
  {\includegraphics[width=5.5cm] {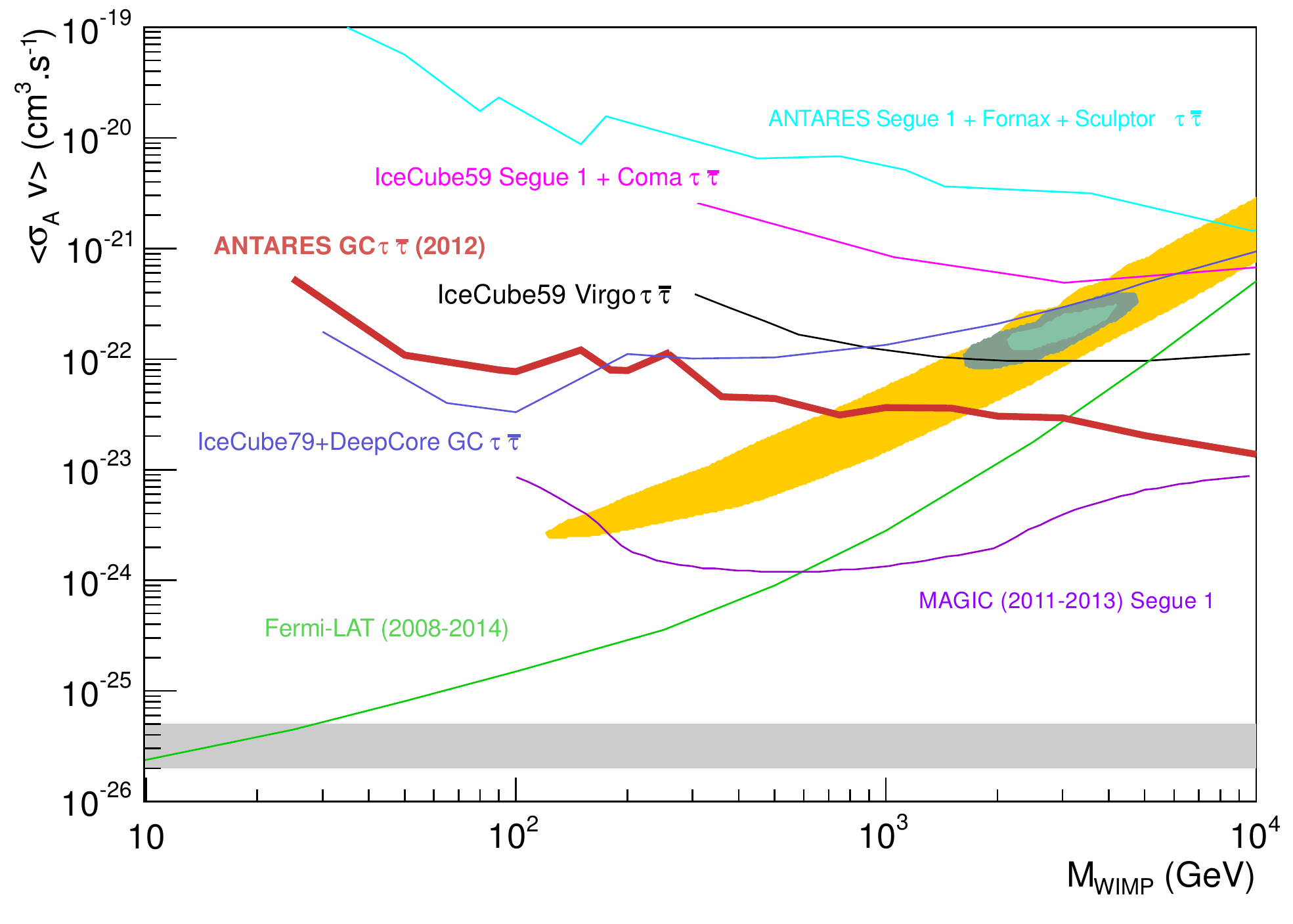} \label{dmsa} }
}
\end{figure}

ANTARES has placed limits on different WIMP dark-matter scenarios by searching for high-energy neutrino emission 
from WIMP annihilation in the Sun, Earth, Galactic Centre, and dwarf galaxies.
Since the dark-matter density is expected to be strongly peaked near the centres of these objects the excellent 
angular resolution of ANTARES, yields competitive limits for WIMP masses above 50 GeV.

Limits on the spin-dependent (WIMP-proton) interaction cross section $\sigma^p_{SD}$ from ANTARES observations of 
the Sun (\fref{dmsd}) \cite{DMSun2016} and on the WIMP-WIMP velocity-averaged self-annihilation cross section  $\sigma_A\nu$ from the Galactic Centre using the $\tau \tau$ channel \cite{GC2015} are given in \fref{dmsa}, and are described in further detail in Ref. \cite{ICRC1207}. 

Dark-matter analyses by ANTARES also includes a search for a WIMP signature from the centre of the Earth 
\cite{ICRC1110} and a test of secluded dark-matter models in the Sun \cite{SDM2016,ICRC1212}.

ANTARES also places limits on beyond-the-standard-model physics, with searches for magnetic monopoles and nuclearites. 
Updates to existing limits are presented in Ref. \cite{ICRC1060}.

\section{Conclusions}

The ANTARES neutrino telescope has proved to be a highly successful instrument,
performing a wide range of physics analyses. In particular, its excellent angular resolution on both
muon-track and cascade events, facilitated by the excellent optical properties of deep-sea water, is well suited
to studying point-like sources of neutrinos. This capability has allowed relevant constraints to be placed upon the 
the origin of the astrophysical neutrino flux reported by IceCube and in particular any possible galactic contribution. 
ANTARES will continue data-taking until end of 2016, thus most of the analyses reported here will be extended with 
an additional three years of data.

%\bibliographystyle{ws-rv-van}
%\bibliography{ws-rv-sample}

%\printindex[aindx]                 % to print author index
%\printindex                         % to print subject index
\end{document}